\documentclass[prl,twocolumn,showpacs,preprintnumbers,amsmath,amssymb]{revtex4}
\usepackage{ulem}
\usepackage{graphicx}
\usepackage{dcolumn}
\usepackage{bm}
\begin{document}
\normalem

\title{Superfluid to Bose-Glass Transition in a 1D Weakly Interacting Bose Gas}

\author{Luca Fontanesi}
\email{luca.fontanesi@epfl.ch}
\affiliation{Institute of Theoretical Physics, Ecole Polytechnique F\'ed\'erale de Lausanne EPFL, CH-1015 Lausanne, Switzerland}
\author{Michiel Wouters}%
\affiliation{Institute of Theoretical Physics, Ecole Polytechnique F\'ed\'erale de Lausanne EPFL, CH-1015 Lausanne, Switzerland}
\author{Vincenzo Savona}
\affiliation{Institute of Theoretical Physics, Ecole Polytechnique F\'ed\'erale de Lausanne EPFL, CH-1015 Lausanne, Switzerland}

\date{\today}

\begin{abstract}
We study the one-dimensional Bose gas in spatially correlated disorder at zero temperature, using an extended density-phase Bogoliubov method. We analyze in particular the decay of the one-body density matrix and the behaviour of the Bogoliubov excitations across the phase boundary. We observe that the transition to the Bose glass phase is marked by a power-law divergence of the density of states at low energy. A measure of the localization length displays a power-law energy dependence in both regions, with the exponent equal to $-1$ at the boundary. We draw the phase diagram of the superfluid-insulator transition in the limit of small interaction strength.
\end{abstract}

\maketitle
Interplay of disorder and many-body interactions is at the origin of a
variety of interesting phenomena. As an example, a particle in a
one-dimensional disorder potential always displays localization~\cite{Kramer}. In a
many-body system, on the other hand, delocalization can arise as a
consequence of interactions~\cite{giamarchi}. A related phenomenon in a repulsive Bose
gas, is the quantum phase transition from a superfluid to an insulating
\emph{Bose glass} phase, as disorder is increased
~\cite{giamarchi,fisher_superfluidinsulator,scalettar}. Trapped
ultracold atomic gases are an ideal system for the investigation of the
Bose-glass phase, thanks to the ability to tune the disorder amplitude
and the interaction strength~\cite{esslinger_weak,hulet,inguscio_boseglass,white}. The
achievement of the superfluid-Bose glass transition is however still
under debate~\cite{zoller,delande}. On the theoretical side, most
efforts have been devoted to the strongly interacting regime of the
phase diagram, through the study of the disordered Bose-Hubbard
model~\cite{reviews,prokofev,rapsch}.
Less attention has been paid to the weakly interacting disordered Bose
gas \cite{lugan_phase,gurarie,zhang,pavloff,lugan_anderson} -- a regime closer to that of recent experimental studies of Anderson
localization in an ultracold Bose
gas~\cite{aspect_anderson,inguscio_anderson}. The weakly interacting
Bose gas in a continuous disorder potential is well described by a mean
field approach, that provides closed-form expressions for the
correlation functions. In low dimensional systems, that are of particular experimental interest \cite{hulet,aspect_anderson,inguscio_anderson}, this approach requires
special care in the description of phase fluctuations, that play a
dominant role by triggering the quantum phase transition. Hence, a
generalization of the theory to quasi-condensates (condensates with
fluctuating phase) is necessary~\cite{mora,andersen,petrov}.

In this Letter, we present a study of the 1D disordered Bose gas within the extended number conserving Bogoliubov theory~\cite{mora}. The main physical quantities under investigation are the spatial coherence, the density of states (DOS) and the inverse participation number (IPN) of the elementary excitations. This latter is a good estimate of the spatial extent of the wave function, although it does not necessarily coincide with an exponential decay length of its tails \cite{Kramer}. We show that all three quantities allow to trace the mean field limit of the phase boundary between the insulating Bose glass and the superfluid quasi-condensed phases. We confirm the scaling of the IPN  with energy, found by Gurarie {\em et al.}~\cite{gurarie} and extend it into the Bose glass phase. Surprisingly the generally accepted fact~\cite{fisher_superfluidinsulator,pavloff,zhang} that the density of states is constant throughout the transition is contradicted by our numerical results: we find that the density of states of the Bogoliubov excitations diverges in the Bose glass phase. Furthermore we show that the loss of spatial coherence in the Bose glass phase is dominated by weak links, across which the coherence drops sharply. Finally, we present numerical evidence that the critical disorder amplitude scales with the interaction energy as a power law.

The N-body Hamiltonian describing the Bose system has the form
\begin{eqnarray}
\hat H  =  \int \mathrm{d}{r} \; \left[\hat{\Psi}^\dagger({r}) \hat H_0  \hat{\Psi}({r}) + \frac{g}{2}\hat{\Psi}^\dagger({r})\hat{\Psi}^\dagger({r})\hat{\Psi}({r})\hat{\Psi}({r})\right],
\end{eqnarray}
where $\hat{H}_0=-\hbar^2\partial_r^2/(2m)+V({r})$ is the noninteracting Hamiltonian, $\hat{\Psi}({r})$ is the field operator, $g$ is the coupling constant  and $V({r})$ is the disorder potential. Here we consider the case of a Gauss-distributed and Gauss-correlated random potential $V({r})$, i.e. such that $\langle V({r})V({r'})\rangle=\Delta^2 e^{-\frac{(r-r')^2}{2\eta^2}}$, where $\Delta$ is the Gauss amplitude and $\eta$ is the spatial correlation length.

In the mean field limit, it is useful to rewrite the annihilation operator in terms of a c-number quasi-condensate density $\rho_0$, and operators for the density and phase fluctuations, $\delta \hat \rho$ and $\hat \theta$ respectively: $\hat\Psi=e^{i\hat\theta}\sqrt{\rho_0+\delta\hat\rho}$. A correct definition of the phase operator, in the extended Bogoliubov method~\cite{mora}, requires the definition of a grid with step size $l$ that fulfills the requirement on the total density $\rho l=(\rho_0+\langle\delta\hat{\rho}\rangle) l> 1$.  It can be shown that the quasi-condensate density obeys the Gross-Pitaevskii equation~\cite{pitaevskii}
\begin{equation}
\left[\hat{H}_0+g\rho_0({r})\right]\sqrt{\rho_0({r})}=\mu\sqrt{\rho_0({r})},
\label{GPE}
\end{equation}
where $\mu$ is the chemical potential. The density and phase fluctuations can be expressed in terms of the usual $u$ and $v$'s of the standard Bogoliubov theory~\cite{shevchenko}, that obey the Bogoliubov-de Gennes equations
\begin{eqnarray}
\left(\hat{H}_0+2g\rho_0({r})-\mu\right)u_j({r}) + g\rho_0({r}) v_j({r}) & = & E_j u_j({r}),\nonumber \\
- g\rho_0({r}) u_j({r}) - \left(\hat{H}_0+2g\rho_0({r})-\mu\right)v_j({r}) & = & E_j v_j({r}).\nonumber \\
\label{BdG}
\end{eqnarray}
The number-conserving formalism requires an orthogonalization of the Bogoliubov modes $u_j({r})$ and $v_j({r})$, with respect to the quasi-condensate density $\rho_0({r})$, thus obtaining the modes $u_{\perp j}({r})$ and $v_{\perp j}({r})$. Using Wick's theorem, the one-body density matrix $G({r},{r'})=\langle\hat\Psi^\dagger({r})\hat\Psi({r'})\rangle$ takes the form~\cite{mora}
\begin{equation}\label{g1}
G({r},{r'}) = \sqrt{\rho({r})\rho({r'})}e^{-\frac{1}{2}\displaystyle\sum_j \left|\frac{v_{\perp j}({r})}{\sqrt{\rho_0({r})}} -\frac{v_{\perp j}({r'})}{\sqrt{\rho_0({r'})}}\right|^2},
\end{equation}
where at $T=0$ only the contribution from the quantum fluctuations appears.
The approximations involved in deriving Eq. (\ref{g1}) require small density  $\delta \hat \rho/\rho_0\ll 1$ and phase fluctuations $\delta \hat \theta/ \rho_0\ll 1$. This mean field description holds in the high-density limit while in the low-density limit the regime of impenetrable bosons is reached and this approach is no more reliable. Throughout this work, we assume a spatially averaged quasi-condensate density $\bar\rho_0\eta=8$  and, for the numerical calculations, we consider a finite system of length $L$ and step size $l$. The simulations that follow are performed for $L=4096\eta$. We define the interaction energy $U$, through the relation $UL|\phi_0({r})|^2=g\rho_0({r})$, with the normalized quasi-condensed wavefunction $\phi_0$. In order for this discretized model to describe the continuous case, we always fulfill the basic requirement that the kinetic hopping energy $t=\frac{\hbar^2}{2ml^2}$ be much larger than any other characteristic energy of the system. This must hold in particular for the energies $U$, $\Delta$, $\mu$ and $E_c=\frac{\hbar^2}{2m\eta^2}$. This latter in turn implies that $\eta\gg l$. In our simulations $t/E_c=16$, namely $\eta=4l$. We solve the GPE using the Crank-Nicholson algorithm for the imaginary-time evolution of $\rho_0$. We then solve the linear Bogoliubov-de-Gennes problem by numerical diagonalization. Periodic boundary conditions are assumed both for the equations and for the randomly generated disorder potential $V({r})$.

We first study the long-range behaviour of the one-body density matrix. In the quasi-condensed phase this quantity is expected to have a power-law decay, as in the spatially uniform gas~\cite{schwartz,mora,andersen}. In presence of disorder, a transition to an exponential decay of the spatial correlation -- characterizing the Bose glass phase -- is expected below a critical value of $U$~\cite{fisher_superfluidinsulator}.  The quantity $G(r,r')$ is affected by the specific shape of the disorder realization. Spatial average leads to the \emph{degree of coherence}
\begin{equation}
g_{1}({r})=\frac{1}{L}\int \mathrm{d}{r'} \frac{G({r},{r'})}
{\sqrt{\rho({r})\rho({r'})}}.
\end{equation}
In our simulations, this quantity reveals to be self-averaging and reproduces directly the decay of the realization-averaged one-body density matrix.
Our numerical analysis shows that the $|v_{\perp j}|^2$ diverge as  $1/E_j$ for $E_j\rightarrow0$ (in what follows, the zero of the energy scale is taken at the chemical potential $\mu$).
This behaviour seems to be generic, because it is also found in two limiting cases that allow for a simple analytical solution: the homogeneous gas~\cite{pitaevskii} and a Josephson junction in the limit of small tunneling~\cite{leggett_josephson} (see below).
Inspection of Eq. (\ref{g1}) then shows that the decay of long range correlations is driven by low energy excitations.
To speed up calculations, we therefore computed only the first $N_{max}=2048$ eigenstates, from which we extracted the one-body density matrix. We checked the convergence of the long range spatial coherence as a function of the cutoff $N_{max}$ and system size $L$.

\begin{figure}
\includegraphics[width=0.5\textwidth]{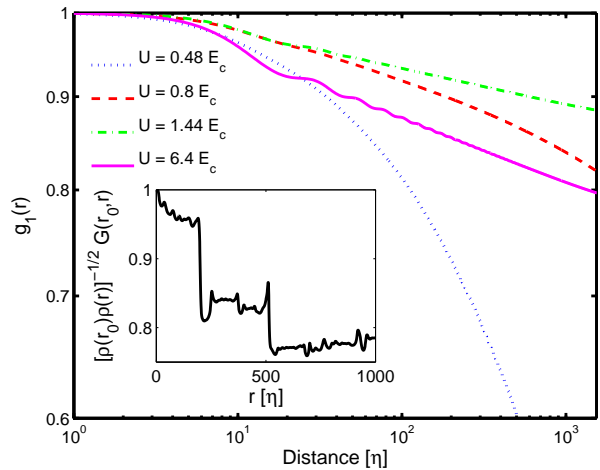}
\caption{\label{fig:g1} $g_1(r)$ in double logarithmic scale, computed for a single realization of the disorder at fixed $\Delta=0.8E_c$, for different $U$.  \emph{Inset}: $G(r,r_0)$ in the Bose glass case ($U=0.48E_c$).}
\end{figure}

Fig. \ref{fig:g1} shows the simulated $g_{1}({r})$ at varying $U$ for fixed $\Delta=0.8E_c$. The inset shows the quantity $G({r}_0,{r})$ computed for $U=0.48E_c$, deep in the Bose-glass phase. We point out that our analysis of the long-range decay of $g_{1}({r})$ is carried out within the range $[0,L/4]$, in order not to be affected by the periodic boundary conditions. The quantity $g_1({r})$ shows an exponential decay for small $U$, characterizing the Bose-glass insulating phase. By increasing $U$, it then increases at fixed $r$, up to a point where the interaction drives the system into a superfluid phase, marked by a power-law decay.
By further increasing $U$, the correlation at fixed $r$ decreases again, in analogy with the decrease of coherence for increasing interactions in the spatially homogeneous case~\cite{pitaevskii}. When increasing $\Delta$ at fixed $U$ instead (not shown), the coherence always shows a monotonic decrease.

The decay of correlations in the Bose glass phase is dominated by low-energy excitations having a phase flip character.  Where a low energy excitation has a node, the coherence drops suddenly with a step that is related to the amplitude of this excitation. This behaviour of $G({r}_0,{r})$ is illustrated in the inset of Fig. \ref{fig:g1}. Physically, the two parts of the Bose gas on each side of the step can be seen as a weakly coupled Josephson junction.
In the superfluid phase, no such abrupt drops in the coherence are found, but rather a smooth behaviour akin to the uniform system.

It is interesting to note that the functional behaviour of the long range spatial coherence is determined by the interaction energy $U$ and not by the density alone.  Indeed, it is determined by the $v$'s and the Bogoliubov-de Gennes equations~(\ref{BdG}) only depend on the product $g \rho_0$. Eq.~(\ref{g1}) shows that the spatial coherence is reduced for decreasing density when keeping the interaction energy constant.

\begin{figure}
\includegraphics[width=0.5\textwidth]{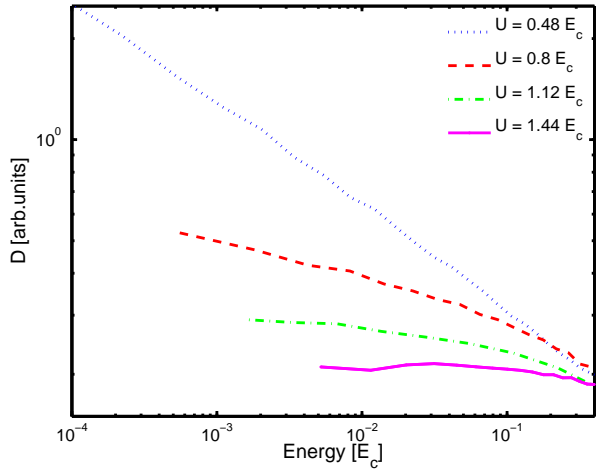}
\caption{\label{fig:DOS}Averaged $D(E)$ at fixed $\Delta=0.8\,E_c$ for various interaction energies.}
\end{figure}

We now turn to study the DOS of the Bogoliubov excitations, defined as $D(E)=\sum_j \delta(E-E_j)$. In the quasi-condensate phase, we expect $D(E)$ to approach a constant for $E\rightarrow0$, similarly to phonons in random elastic chains~\cite{ziman}. For the Bose-glass phase, it has been argued~\cite{fisher_superfluidinsulator,zhang} that the low-energy limit of the DOS should remain constant. Our results, however, do not support this behaviour in the mean field limit. We show in Fig. \ref{fig:DOS}, the quantity $D(E)$ plotted for 4 values of $U$. For the largest $U$ it clearly displays a constant limiting value for $E\rightarrow0$, while $D(E)$ develops a power-law divergence for the smallest $U$. The numerical accuracy of the data makes it difficult to extract the limiting behaviour of $D(E)$ for intermediate values of $U$, although the power-law divergence in the Bose-glass phase is clearly assessed.
We propose the following handwaving argument in support of the power-law divergence of the DOS in the glass phase. By inspecting Eq.(\ref{g1}) for fixed $r'=r_0$, we infer that the asymptotic behaviour at large $r-r_0$ is mainly determined by the term $\sum_j|v_{j\perp}(r)|^2$ in the exponent. Thus, an estimate of Eq.(\ref{g1}) in that limit is $G(r,r_0)\sim\exp[\int |v_{E\perp}(r)|^2 D(E) \mathrm{d}E]$ (see also Ref.~\cite{zhang}). Since $|v_{E\perp}(r)|^2$ always shows a $1/E$ divergence for $E\rightarrow0$, then the change of $G(r,r_0)$ from power-law to exponential must be determined by $D(E)$ passing from constant to a power-law divergence.

Another quantity of interest in connection with recent experiments~\cite{aspect_anderson,inguscio_anderson} is the localization length of the Bogoliubov modes~\cite{pavloff,lugan_anderson,gurarie}. For a single particle, all wave functions are expected to be exponentially localized in 1D~\cite{thouless}, with the IPN monotonically increasing from a finite value at $E\rightarrow-\infty$. In presence of interactions, as already pointed out, delocalized low-energy \emph{phase}-excitations represent the major mechanism of long-range decoherence. The behaviour at $E\rightarrow0$ in the quasi-condensate phase should diverge as a power law $E^{-\alpha}$ but the value of $\alpha$ is still object of controversy~\cite{pavloff,gurarie}. To study the localization of the excitations, we have computed their inverse participation number \begin{equation}
\frac{1}{I_j} = \frac{\int \mathrm{d}{r} |v_{\perp j}({r})|^4}{\left(\int \mathrm{d}{r}|v_{\perp j}({r})|^2\right)^2},
\end{equation}
and the corresponding realization-averaged quantity $L(E)=\sum_j I_j \delta(E-E_j)/D(E)$. We focus here on the Bogoliubov $v_\perp$ modes that are the only contribution to decoherence  at $T=0$.
\begin{figure}
\includegraphics[width=0.5\textwidth]{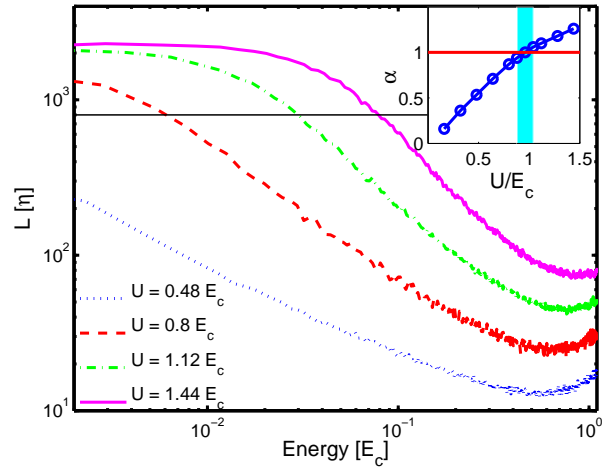}
\caption{\label{fig:ll} Averaged $L(E)$ for varying interaction at fixed $\Delta=0.8\,E_c$. The horizontal line indicates a limit where the finite size effects start to play a role leading to saturation of $L(E)$. \emph{Inset}: Exponent of the power-law divergence of  $L(E)$ as a function of $U/E_c$: the shaded zone marks the phase transition.}
\end{figure}
Our results for $L(E)$ are displayed in Fig. \ref{fig:ll}. Remarkably, we obtain a power-law divergence $E^{-\alpha}$ for $E\rightarrow0$, independently of the gas phase. The exponent $\alpha$ (plotted in the Inset of Fig. \ref{fig:ll}) varies continuously from $\alpha\ll1$ for the lowest value of $U$, to $\alpha>1$ for the largest $U$ considered.
The finite size of the simulations limits the analysis for large $U$, deep in the quasi-condensate phase. This supports the scenario recently proposed by Gurarie {\em et al.}~\cite{gurarie}, where $\alpha$ equals unity at the phase boundary and increases to 2 for increasing interaction strength. We thus tentatively identify the $\alpha=1$ case as the phase boundary. Moreover we find that $\alpha$ continuously decreases when going deeper in the insulator phase, apparently linearly vanishing for $U\rightarrow0$. This picture is consistent with the constant limiting value of $L(E)$ at low energy, expected for the non-interacting case.

\begin{figure}
\includegraphics[width=0.5\textwidth]{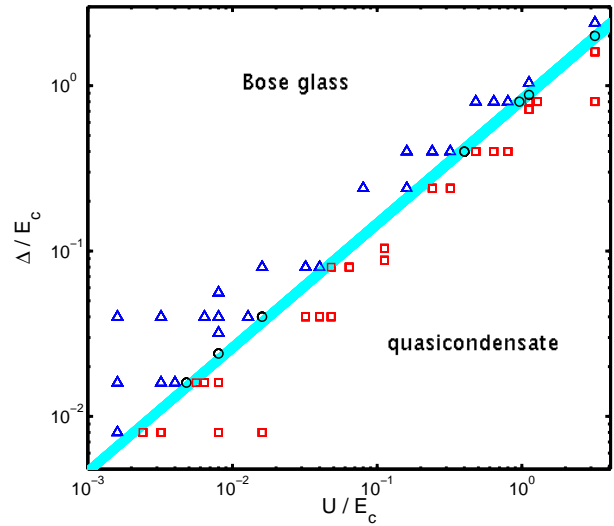}
\caption{Sketch of the phase diagram of the 1D Bose gas as a function of interaction and disorder. ($\triangle$): Bose glass; ($\square$): quasi-condensate. The points denoted by a circle are those for which an asymptotic behaviour cannot be extracted from our simulations, implying proximity to the phase boundary. These points are therefore included in the shaded region that marks the transition.}
\label{fig:phase}
\end{figure}

The three quantities studied above, display a phase boundary for the same values of $\Delta$ and $U$, within the numerical accuracy. This allows to draw, in Fig. \ref{fig:phase}, a phase diagram for the quasi-condensate to Bose-glass transition at zero temperature close to the origin of the $U-\Delta$ plane. Symbols denote simulated systems in both phases. The shaded region contains the phase boundary and its width denotes the uncertainty in extracting an asymptotic behaviour from the finite system size. This uncertainty was reduced by the joined analysis of $g_1(r)$, the DOS and the IPN. In particular, it is remarkable how the boundary coincides with the $\alpha=1$ exponent for the IPN. The boundary appears to obey a power law $\Delta/E_c=C(U/E_c)^\gamma $, with an exponent $\gamma=0.75\pm0.03$. Finally, we remind that, at densities much lower than here considered, the mean field approximation becomes invalid. In the Bose glass phase, this occurs when the coherence is limited to a single maximum of the quasi-condensate density, the so-called Lifshitz glass phase~\cite{lugan_phase}. In the superfluid phase, the Bogoliubov approximation breaks down when the coherence is strongly reduced on the scale of the interparticle distance and an interaction dominated Bose glass phase is reached. The resulting reentrant behaviour of the glass phase as a function of interaction strength~\cite{scalettar,prokofev} can consequently not be described within Bogoliubov theory.

In conclusion, we have derived the mean field limit of the phase diagram for a 1D Bose gas in presence of continuous, spatially correlated disorder. While analogous studies exist on the Bose-Hubbard lattice model, not much attention had been devoted to this limiting case. The first experimental works on the dilute Bose gas in a disorder potential~\cite{aspect_anderson,inguscio_anderson} essentially considered the case of vanishing interactions. In relation to the phase diagram, we have estimated for both these experiments the value $U/E_c\sim1$. In Ref. \cite{aspect_anderson} $\Delta/E_c\sim0.05$. In Ref. \cite{inguscio_anderson} the disorder parameter $\Delta/J$, corresponding to our $\Delta/t$, is varied over a broad range. The continuous limit here considered is only appropriate to the situation $\Delta/t\ll1$. The results here obtained provide useful indications for future experiments aimed at the characterization of the Bose glass phase boundary.

We are grateful to I.~Carusotto, D.~Sarchi and P.~Lugan for enlightening discussions. This work was supported by the Swiss National Science Foundation through project No. 200021-117919.

\end{document}